\documentclass[notitlepage,superscriptaddress,showpacs,nobalancelastpage,twocolumn,aps,prl]{revtex4-1}
\usepackage[english]{babel}
\RequirePackage[T1]{fontenc}
\RequirePackage{times} 

\usepackage{amsfonts}
\usepackage{subfigure}
\usepackage{amsmath}
\usepackage{amssymb}
\usepackage{graphicx,epstopdf}
\usepackage{array}
\usepackage{color}
\usepackage{my_symbols}
\usepackage{CJK}
\usepackage{bm}
\usepackage[normalem]{ulem}

{\par}

\newcommand\rmv{\bgroup\markoverwith {\textcolor{red}{\rule[0.5ex]{2pt}{0.4pt}}}\ULon}

\begin{document}
\begin{CJK*}{UTF8}{gbsn} 
\title{Spin-Wave Fiber}
\author{Weichao Yu (余伟超)}
\thanks{These authors contributed equally.}
\affiliation{Department of Physics and State Key Laboratory of Surface Physics, Fudan University, Shanghai 200433, China}
\author{Jin Lan (兰金)}
\thanks{These authors contributed equally.}
\affiliation{Department of Physics and State Key Laboratory of Surface Physics, Fudan University, Shanghai 200433, China}
\author{Ruqian Wu}
\affiliation{Department of Physics and State Key Laboratory of Surface Physics, Fudan University, Shanghai 200433, China}
\affiliation{Department of Physics and Astronomy, University of California, Irvine, California 92697-4575, USA}
\author{Jiang Xiao (萧江)}
\email[Corresponding author:~]{xiaojiang@fudan.edu.cn}
\affiliation{Department of Physics and State Key Laboratory of Surface Physics, Fudan University, Shanghai 200433, China}
\affiliation{Collaborative Innovation Center of Advanced Microstructures, Fudan University, Shanghai 200433, China}

\begin{abstract}
Spin waves are collective excitations propagating in the magnetic medium with ordered magnetizations. Magnonics, utilizing the spin wave (magnon) as information carrier, is a promising candidate for low-dissipation computation and communication technologies. We discover that, due to the Dzyaloshinskii-Moriya interaction, the scattering behavior of spin wave at a magnetic domain wall follows a generalized Snell's law, where two magnetic domains work as two different mediums. Similar to optical total reflection that occurs at the water-air interfaces, spin waves may experience total reflection at magnetic domain walls when their incident angle larger than a critical value. We design a spin wave fiber using a magnetic domain structure with two domain walls, and demonstrate that such a spin wave fiber can transmit spin waves over long distance by total internal reflections, in analogy to an optical fiber. Our design of spin wave fiber opens up new possibilities in pure magnetic information processing.
\end{abstract}

\maketitle
\end{CJK*}

A trend in post-silicon information processing is the development of systems that employ (quasi-) particles other than electrons as the information carriers to avoid the the Joule heating. One of the most promising candidates is magnonics \cite{kruglyak_magnonics_2010,chumak_magnon_2015}, which transfers information through \rmv{the} collective excitations of magnetization, or called spin waves (magnons).The crucial advantage of using spin waves  as information carriers lies in that they can propagate in both conducting and insulating magnetic materials without physical motion of electrons. This enables the development of insulator-based information systems that produce no Joule heating at all. In addition, spin waves can be manipulated via magnetic structures in a single material instead of heterogenous structures of several different materials. This makes it possible to realize a rewritable spin wave logic architecture \cite{lan_spin-wave_2015}. Due to these desirable features, magnonics becomes a new realm of active interdisciplinary research and we have witnessed rapid and fruitful developments in recent years. \cite{kajiwara_transmission_2010,vogt_realization_2014,chumak_magnon_2014,lan_spin-wave_2015,garcia-sanchez_narrow_2015,cheng_antiferromagnetic_2015}. A key issue for the development of magnonic devices is the design of ``wires'' that guide the propagation of spin waves (magnons).

The World Wide Web (Internet), one of the most important infrastructures of the modern society, would not be possible without the optical fiber, which transmits information across long distance with little loss.
Here we report a design of the spin wave fiber that carries spin waves with little loss, in an analog of the optical fiber  but transmits spin waves rather than photons. The design is based on a magnetic domain structure with two domain walls, at which spin waves are totally reflected due to the Dzyaloshinskii-Moriya interaction (DMI) \cite{dzyaloshinsky_thermodynamic_1958,moriya_anisotropic_1960}, an asymmetric magnetic interaction induced by the spin-orbit interaction. An optical fiber is a cylindrical structure consisting of a core and a cladding layer, which are made of different dielectric materials. The different refraction indices of these two materials cause the total reflection at the boundary of the core and cladding layers \cite{senior_optical_2009}. In contrast, it is feasible and convenient to use one single material but different magnetic domains for the core and cladding layer in a spin wave fiber. Typically, the spin wave dispersion relation does not vary much across magnetic domains. However, the DMI leads to a domain-dependent spin wave dispersion, which gives rise to a total reflection of spin waves at the domain walls as needed for the design of spin wave fibers.

\begin{figure}[t]
\centering
\includegraphics[width=0.45\textwidth]{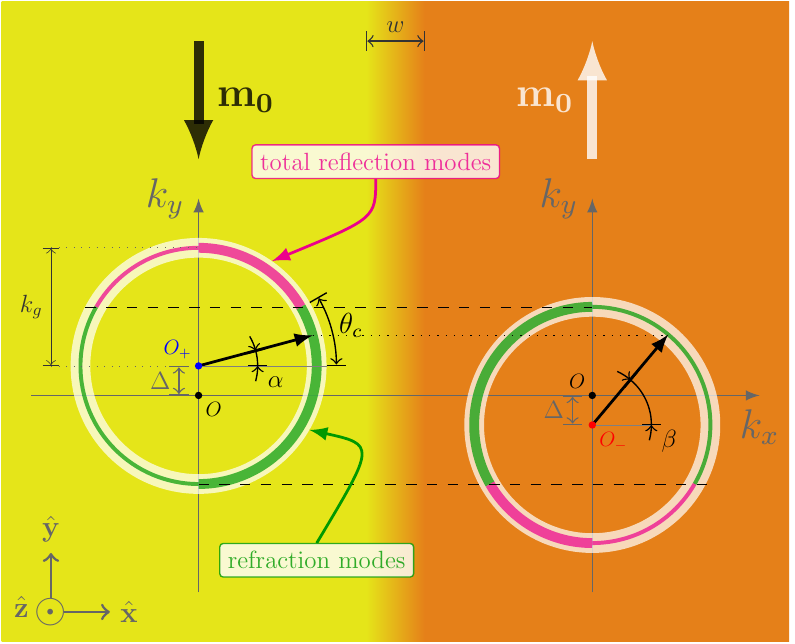}
\caption{Schematic of the magnonic Snell's law. A domain wall structure with magnetization pointing in $\mb_0 = \mp\hby$ in the left/right domain and  $\mb_0$ continuously rotates from $-\hby$ to $+\hby$ over a  characteristic  domain wall width $w$ as depicted in the upper right inset. The isofrequency circles in the wavevector space are shifted for two magnetic domains with opposite magnetizations when DMI is present. The spin wave modes on both circles have the same frequency $\omega$. The modes on the thick green segment can be refracted into the other domain because there exist propagating modes with the same $k_y$ on the isofrequency circle across the domain wall. However, the modes on the thick magenta segment cannot pass and must be totally reflected due to the lack of the propagating modes with the same $k_y$ across the domain wall.  }
\label{fig:tot_refl}
\end{figure}

{\it Model.}
To illustrate the scattering behavior of spin waves at magnetic domain walls, we consider a magnetic thin film with two magnetic domains with opposite magnetization directions, and a Bloch domain wall in between as shown in Fig. \ref{fig:tot_refl}. The film is in the $x$-$y$ plane and has an in-plane easy-axis along $\hby$. The magnetization in the left/right domain is along $\mp \hby$ direction, respectively. The domain wall is also along the $\hby$ direction.

In continuum approximation, the magnetic dynamics is described by the Landau-Lifshitz-Gilbert (LLG) equation,
\begin{align}
    {\partial\mb\ov\partial t} =& -\gamma \mb \times \bH_{\rm eff} + \alpha_{\text{\tiny G}} \mb\times {\partial\mb\ov\partial t},
\label{eqn:LLG}
\end{align}
where $\mb(\br,t)$ is the unit vector representing the magnetization direction, $\gamma$ is the gyromagnetic ratio, $\alpha_{\text{\tiny G}}$ is the Gilbert damping parameter, and $\gamma \bH_{\rm eff} = K m_y \hby + A \nabla^2\mb - D\nabla \times \mb$ is the effective magnetic field with $K$ the magnetic uniaxial anisotropy (along the $\hby$ direction), $A$ the exchange coupling constant, and $D$ the parameter for the DMI. Different from the exchange coupling which favors parallel alignment of magnetizations, the DMI tends to promote  helical magnetic structures \cite{bak_theory_1980}. However, if the DMI is not too strong ($D<2\sqrt{AK}$), the exchange coupling still dominates and \rmv{the} magnetization in each domain stays parallel to $\hby$. Let $\mb_0$ be the static magnetization direction, and $\delta\mb=m_\theta \hbe_\theta+m_\phi \hbe_\phi$ be the dynamical excitation on top of the static $\mb_0$, with $\hbe_{\theta,\phi}\perp \mb_0$ as the two transverse directions to $\mb_0$. In the small excitation approximation ($m_{\theta, \phi} \ll 1$), the dispersion relation of the spin wave excitation in the left/right domain with $\mb_0=\mp \hby$ is given by \cite{moon_spin-wave_2013,garcia_nonreciprocal_2014}
\begin{equation}
    \omega(\bk) = K + A (k_x^2 +k_y^2) \mp D k_y,
\label{eqn:dispersion}
\end{equation}
where $\bk=(k_x,k_y)$ is the wavevector for the spin wave mode. In the absence of DMI ($D = 0$), the spin wave modes with frequency $\omega$ form a circle centered at the origin in the wavevector $\bk$-space. However, the DMI, in the last term in \Eq{eqn:dispersion}, pushes the isofrequency circle by $\Delta = D/2A$ in the direction of $-\mb_0$ as depicted separately  for the two domains in \Figure{fig:tot_refl}. The spin wave group velocity calculated from the dispersion is $\bv_g = \partial \omega/\partial \bk= 2A (\bk\mp\hby\Delta) \equiv 2A\bk_g^\mp$,
where $\abs{\bk_g^\mp} = k_g(\omega) = \sqrt{(\omega-K)/A+\Delta^2}$ is the distance from $\bk$ to the shifted origin $O_\pm$, or the radius of the isofrequency circle.

{\it Snell's law for spin wave. } An important consequence of the opposite shifts of the isofrequency circles in the left and right domains is spin wave refraction across the domain wall, when the spin wave strikes from left domain with an incident angle $\alpha$ (angle of the group velocity with respect to the domain wall normal $+\hbx$ direction), it is refracted to a mode with an outgoing angle $\beta$ in the right domain. To guarantee continuity of the wavevector ($k_y$) along the domain wall, the refraction angle obeys the generalized Snell's law \cite{yu_light_2011,yu_flat_2014},
\begin{align}
    k_g \sin\alpha + \Delta = k_g \sin \beta - \Delta.
\label{eqn:snell}
\end{align}
This is an analogy of the generalized Snell's law in photonics \cite{yu_light_2011,yu_flat_2014,sun_gradient_2012,sun_high_2012,chen_dual-polarity_2012} and phononics \cite{li_experimental_2014,zhu_acoustic_2015} across meta-surfaces, in which a phase gradient is introduced at the interface between two mediums by sub-wavelength engineering. In comparison to the complicated meta-surfaces in its optical and acoustic counterparts, it is much easier and more straightforward to realize the non-trivial Snell's law in magnonics, {\it i.e.}, simply by introducing DMI in the whole film without any sub-wavelength engineering. The generalized magnonic Snell's law can lead to the anomalous negative refraction of spin wave \cite{vashkovskii_negative_2004,lock_properties_2008,kim_negative_2008}. Note that in \Figure{fig:tot_refl} the static magnetization in the left/right domain $\mb_0=\mp\hby$ is parallel to the domain wall\rm{, which is also along $\hby$}. In a general situation where  $\mb_0$ in the left/right domain forms an angle $\pm\xi$ with the domain wall (along $\hby$), the offset of the isofrequency circle along $\hby$ is reduced to $\Delta\cos\xi$, thus the magnonic Snell's law in \Eq{eqn:snell} is modified with a substitution $\Delta\ra\Delta\cos\xi$.

According to the magnonic Snell's law in Eq. \eqref{eqn:snell}, total reflection occurs when the incident angle $\alpha$ satisfies $\theta_c < \alpha < \pi/2$ (the thick magenta segment in \Figure{fig:tot_refl}) with the critical incident angle (provided $k_g(\omega) > 2\abs{\Delta}$)
\begin{equation}
\theta_c =  \arcsin\midb{1-{2\Delta\ov k_g(\omega)}}.
\label{eqn:theta_c}
\end{equation}
As a result, the modes with incident angle $-\pi/2<\alpha<\theta_c$ (the thick green segment in \Figure{fig:tot_refl}) are refracted into the right domain. Depending on value of $\Delta/k_g(\omega)$ (thus on $D$ and $\omega$), the critical angle $\theta_c$ can take any value between $-\pi/2$ and $\pi/2$. For instance, when there is no DMI ($D = 0$), $\theta_c = \pi/2$ and all incident modes will be transmitted and no total reflection occurs. While when $D$ is strong (or $\omega$ is small) such that the offset between the two isofrequency circles $2\Delta > k_g(\omega)$, $\theta_c = -\pi/2$ and all incident modes will be totally reflected.

\begin{figure*}[t]
\includegraphics[width=0.162\textwidth]{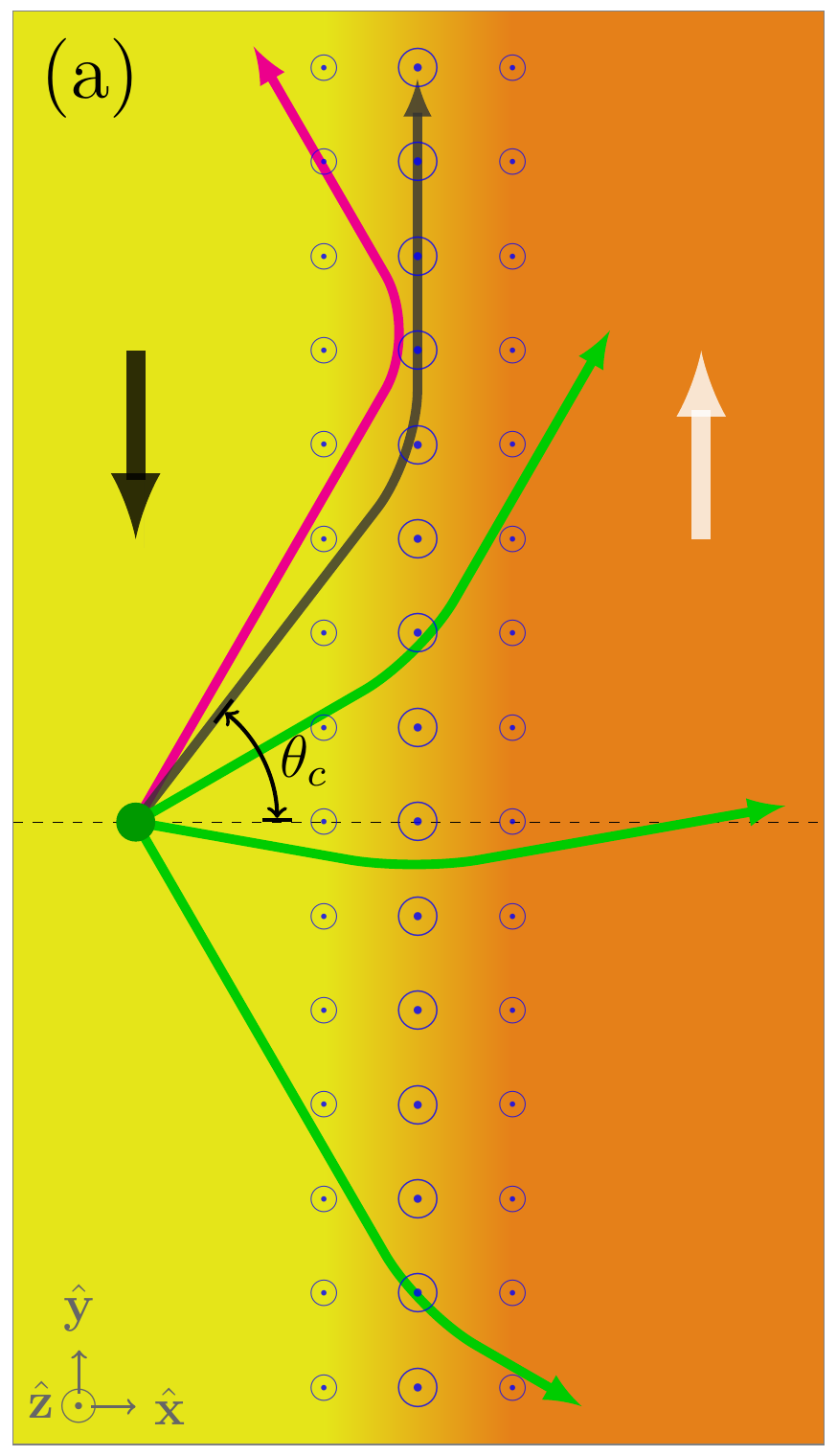}
\includegraphics[width=0.162\textwidth]{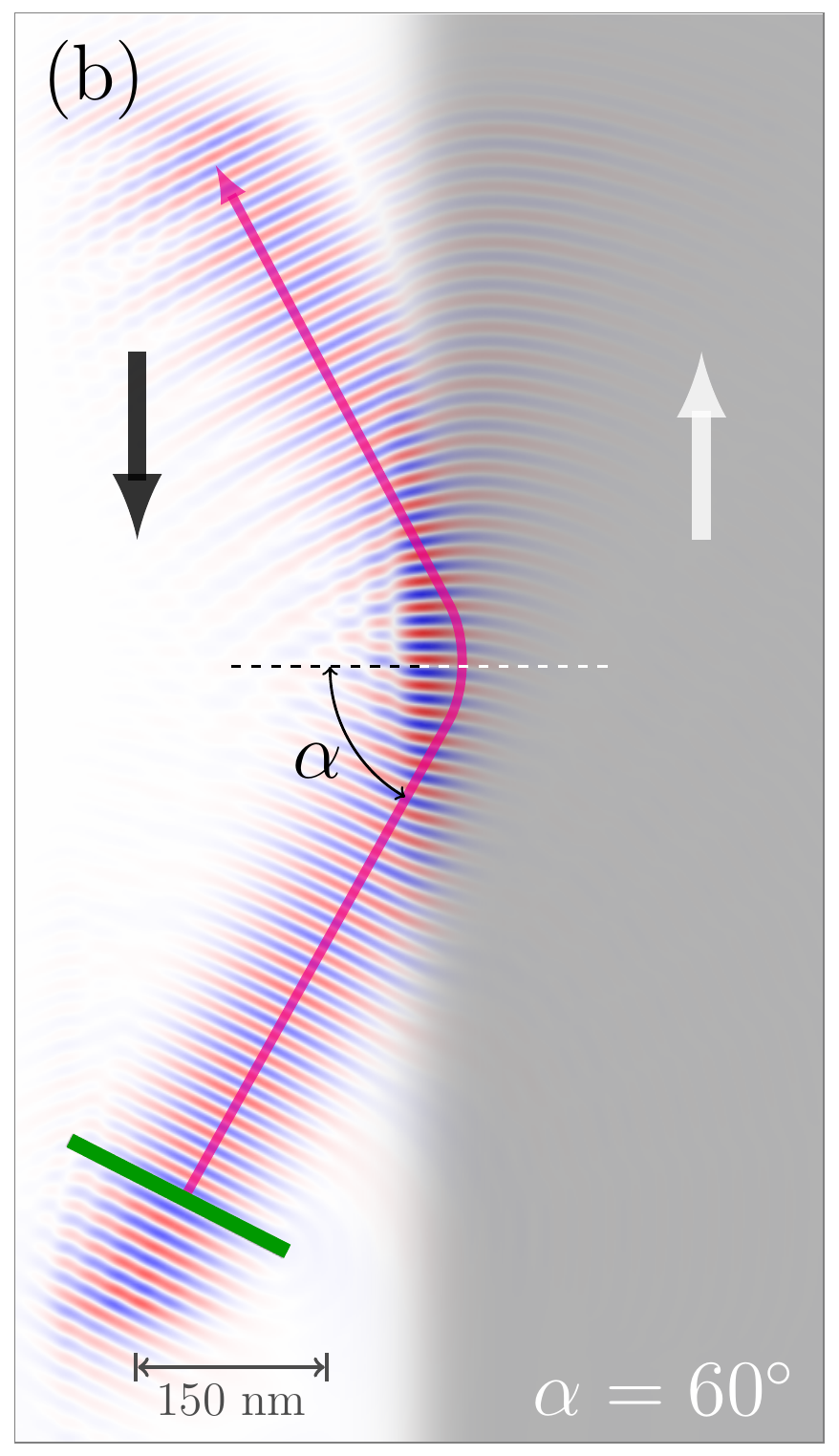}
\includegraphics[width=0.162\textwidth]{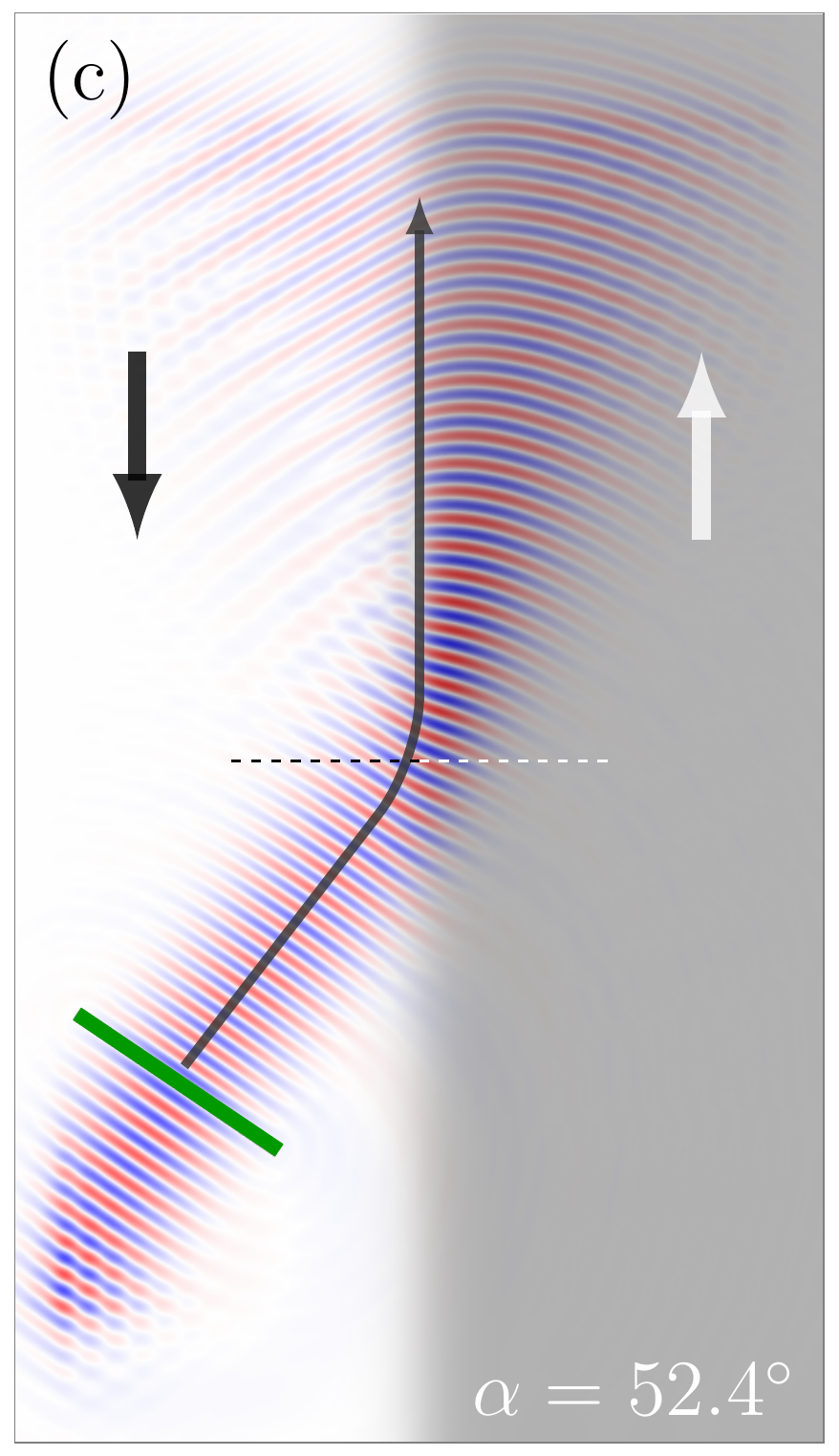}
\includegraphics[width=0.162\textwidth]{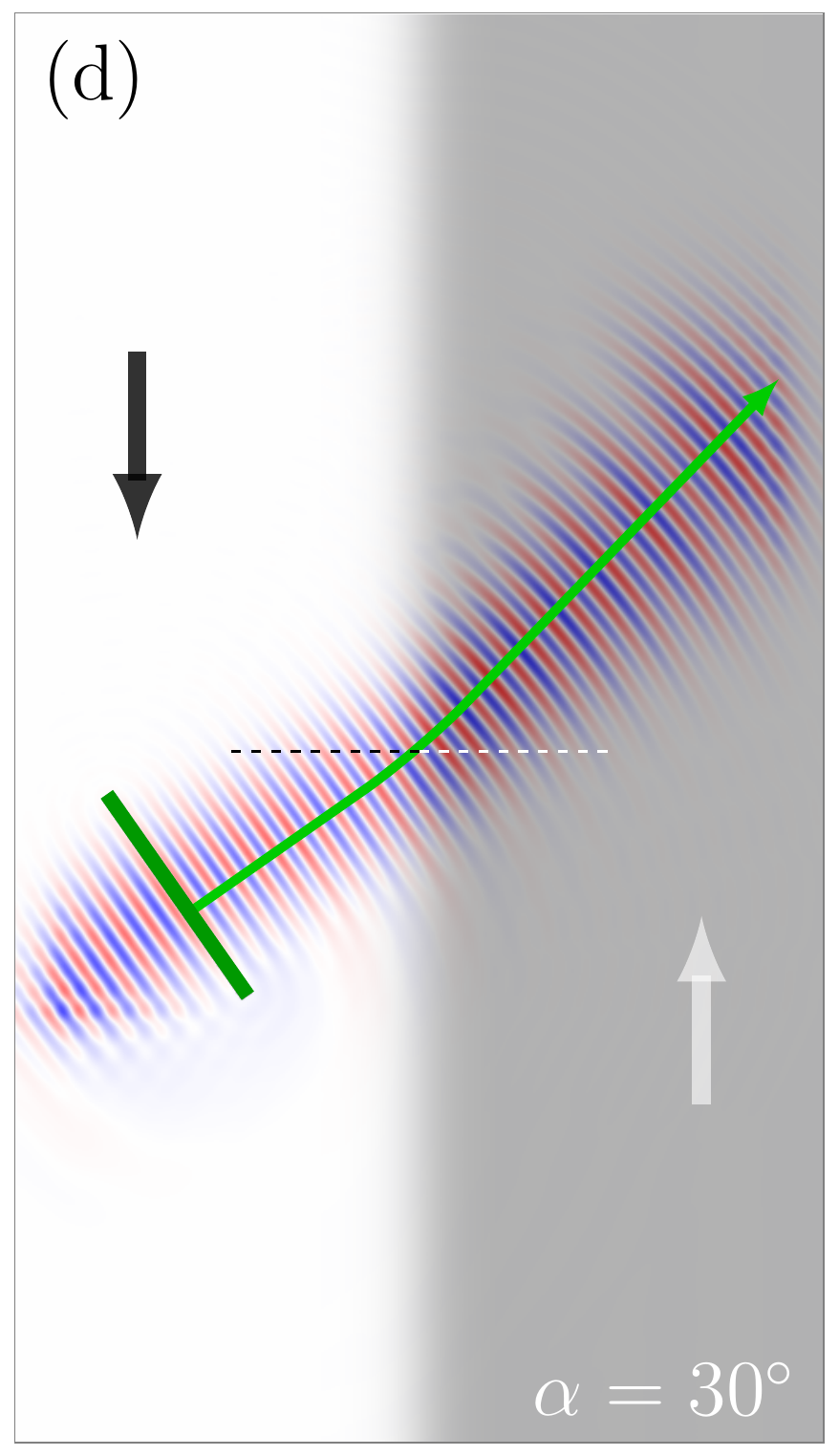}
\includegraphics[width=0.162\textwidth]{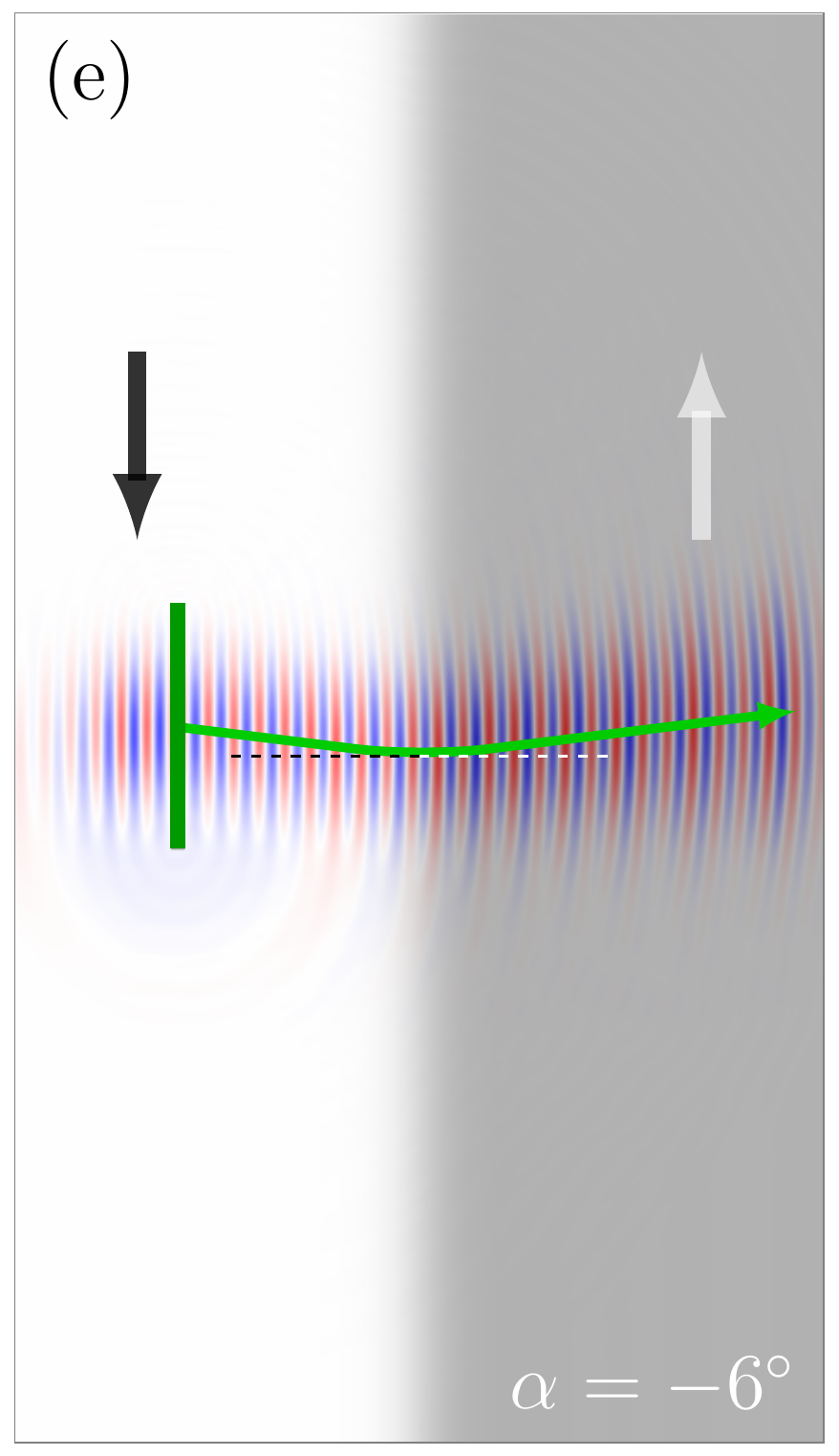}
\includegraphics[width=0.162\textwidth]{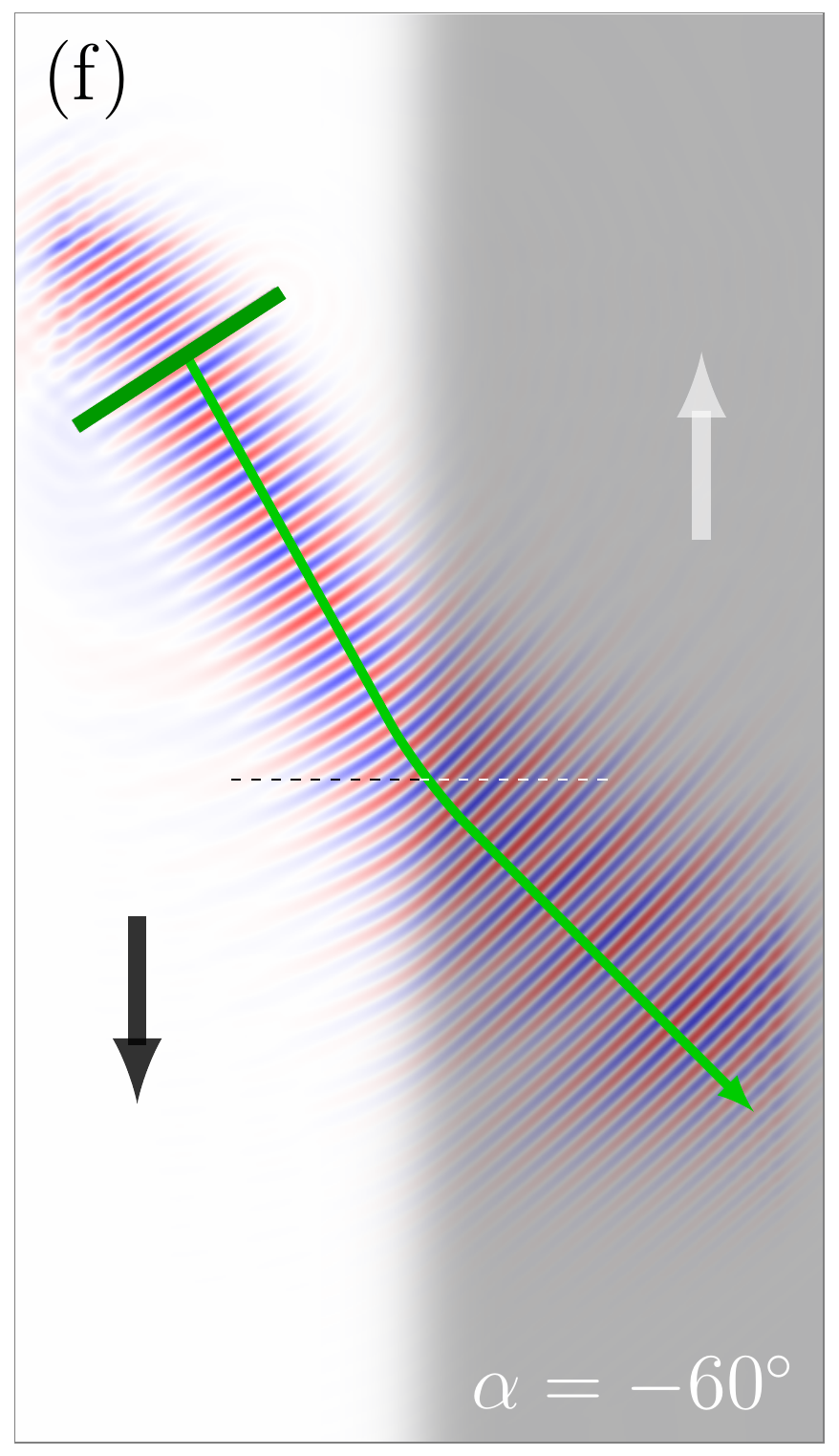}
\caption{Schematic diagram (a) and the magnetic simulations (b-f) for spin wave reflection and refraction at a magnetic domain wall. The black/white arrow denotes the magnetization direction in the left/right domain colored by yellow/orange (white/gray) in the schematic diagram (simulations). (a) Schematic diagram of reflection/refraction behavior for spin wave incident from left domain with various incident angles. Magenta: $\alpha > \theta_c$, total reflection; black: $\alpha = \theta_c$, critical case ; green: $\alpha < \theta_c$, refraction. All trajectories are bent counter-clockwise by the effective magnetic field perpendicular to plane (denoted by the symbol $\odot$). (b-f) Magnetic simulations with the incident angle $\alpha=60^{\circ}$ (b), $\alpha=\theta_c$ (c), $\alpha=30^{\circ}$ (d), $\alpha=-6^{\circ}$ (e), $\alpha=-60^{\circ}$ (f). In all panels: the domain wall width $w\sim 30$nm; the green bar denotes the exciting location of the spin wave; the exciting fields frequency is $f= 100\mathrm{Ghz}$, and the critical angle is estimated to be $\theta_c=52.4^{\circ}$.  }
\label{fig:ref_mag_sim}
\end{figure*}

{\it Semiclassical picture.}
An interesting question to ask is  what exactly happens as the spin wave transmits through the domain wall, that has a finite width $w$, over which the magnetization vector $\mb_0(\br)$ continuously varies from one direction to the other. The magnetic texture of a domain wall can be described by $\mb_0(x)= (\sin\theta_0\cos\phi_0, \sin\theta_0\sin\phi_0, \cos\theta_0)$, where $\theta_0(x),\phi_0(x)$ are the polar and azimuthal angle of $\mb_0$ with respect to $\hbz$. For a Bloch type domain wall, in the presence of DMI, a Walker configuration with $\phi_0(x)=\pi/2$, $\theta_0(x)=-\pi/2-2\mathrm{sign}(D)\arctan[\exp(x/w)]$ is still stable,
where $w=\sqrt{A/K}$ is the characteristic domain wall width. In this configuration, the magnetization continuously rotates from $-\hby$ to $+\hby$ in the $y$-$z$ plane, and the rotation direction is controlled by the sign of DMI constant $D$.

By rewriting the dynamical excitation $\delta\mb$ as $\psi= m_\theta -i m_\phi$, the LLG equation \Eq{eqn:LLG} for $\delta\mb$ can be recast to a Schr\"{o}dinger-like equation for $\psi$ \cite{yan_all-magnonic_2011, wang_magnon-driven_2015,lan_spin-wave_2015}:
\begin{equation}
i\hbar{\partial\psi\ov\partial t} = \midb{ {1\ov 2m^*}\smlb{\hbp-{e\ov c}\bA}^2+V}\psi
\label{eqn:sch},
\end{equation}
where $e$ is the electric charge, $m^*=\hbar/2A$ is the effective mass, $\hbp=-i\hbar\nabla$ is the momentum operator, $V(x) = K[1-2\mathrm{sech}^2(x/w)]$ is the texture-induced effective scalar potential, and $\bA(x) = -(Dm^*e/c) \tanh(x/w) \hby$ is the DMI-induced effective vector potential. The scalar potential $V$ is special because it is reflectionless when there is no DMI. \cite{yan_all-magnonic_2011} The vector potential gives rise to an effective magnetic field $\bB(x) =\nabla \times \bA = -(Dm^*e/ cw) \mathrm{sech}^2(x/w) \hbz$, which is perpendicular to the film plane and maximizes at the center of the domain wall. As $x/w\ra \pm\infty$, the vector potential $\bA(x) \ra \pm D/2A$, which corresponds to the shift of $k_y$ mentioned above in the left/right domain as in \Eq{eqn:dispersion}. With \Eq{eqn:sch}, the scattering behavior for spin waves by a domain wall is exactly the same as that for an electron by a scalar potential $V$ and a vector potential $\bA$ (and the associated field $\bB$).

As mentioned above the scalar potential $V$ is a reflectionless one, thus the spin wave scattering behavior is mostly dominated by the vector potential (and the effective magnetic field). The effective magnetic field $\bB(x)\|\hbz$ is perpendicular to the film, and only exists in the domain wall region, as denoted by the symbol $\odot$ in \Figure{fig:ref_mag_sim}(a). The spin wave scattering behaviors shown in \Figure{fig:ref_mag_sim} can all be understood by considering the effective Lorentz force exerted by this effective magnetic field on an electron in the domain wall region. Evidently, spin waves striking into such an effective magnetic field region will be bent counter-clockwise due to the effective Lorentz force. Consequently, the spin wave that passes through the magnetic field region (the domain wall region), is refracted in counter-clockwise direction. Moreover, if the incident angle is too shallow ($\alpha > \theta_c$), the Lorentz force is able to bend the trajectory so much such that the spin wave is completely turned back, and a total reflection occurs.


{\it Total reflection for micromagnetic simulations.} The refraction and reflection behaviors for spin waves at a domain wall discussed above are confirmed by micromagnetic simulations. According to \Eq{eqn:snell}, the incident and the refracted angles satisfy $-\pi/2\le \alpha < \beta \le \pi/2$ (for $D> 0$), thus the spin wave trajectory should always bend counter-clockwise. \Figure{fig:ref_mag_sim}(a) shows the schematic diagram of the spin wave trajectories for various incident angles: when $\alpha>\theta_c$, the spin wave is totally reflected (magenta trajectory); when $\alpha< \theta_c$, the spin wave is refracted (green trajectories); the black trajectory corresponds to the critical situation with $\alpha = \theta_c$. \Figure{fig:ref_mag_sim}(b-f) show the micromagnetic simulation for the five different incident angles: $\alpha = 60^{\circ}, 52.4^{\circ},30^{\circ},-6^{\circ},-60^{\circ}$, corresponding to the total reflection (b), critical incidence (c), and refraction (d,e,f). Because of the reflectionlessness of the scalar potential $V(x)$ for a Bloch domain wall, the spin wave refractions as in \Figure{fig:ref_mag_sim}(d-f) is not accompanied by any reflection, which is quite different from its optical analog. More interestingly, \Figure{fig:ref_mag_sim}(e) shows the anomalous negative refraction, {\it i.e.}, both incident and refracted trajectories lie in the same side of normal direction of the scattering plane (the domain wall). All spin wave beams in \Figure{fig:ref_mag_sim}(b-f) are bent counter clockwise, as predicted by the magnonic Snell's law \Eq{eqn:snell}. Such bending trajectories can be alternatively understood by semiclassical picture above using the effective Lorentz force acting on the spin wave by the DMI induced effective magnetic field (indicated by symbol $\odot$ in \Figure{fig:ref_mag_sim}(a)) localized at the domain wall region \cite{lan_spin-wave_2015}.

\begin{figure*}[t]
\includegraphics[angle=-90,width=0.49\textwidth]{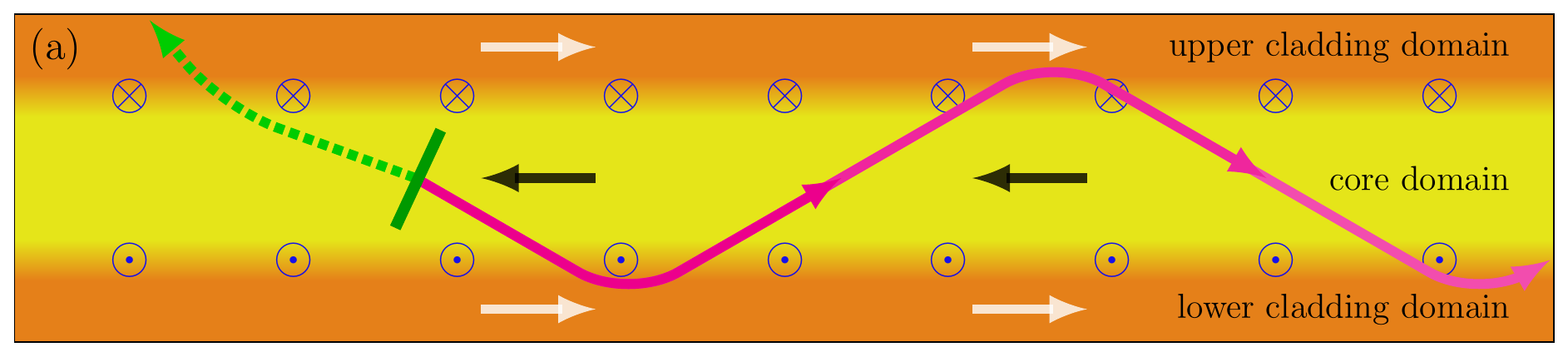}\hspace{0.1cm}
\includegraphics[angle=-90,width=0.49\textwidth]{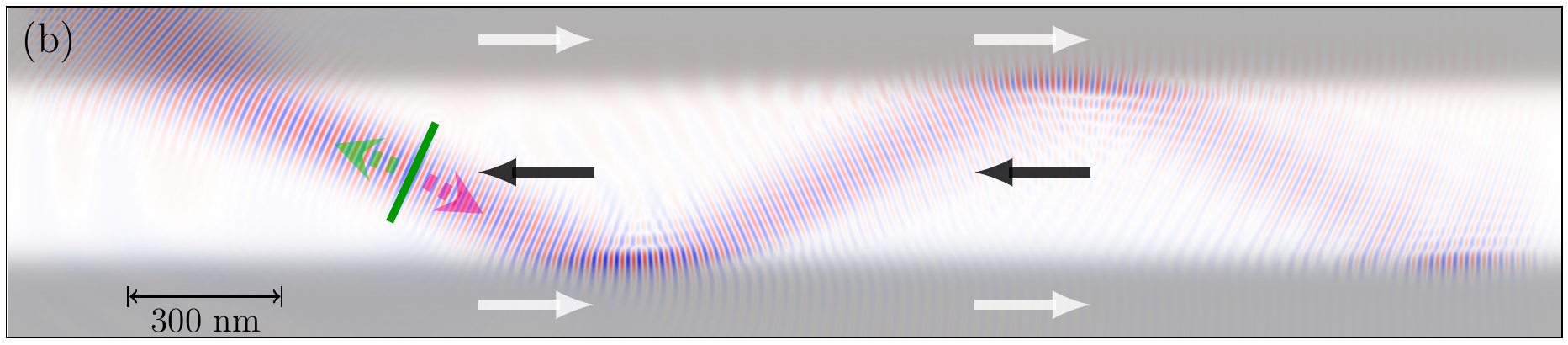}\vspace{0.1cm}
\includegraphics[angle=-90,width=0.995\textwidth]{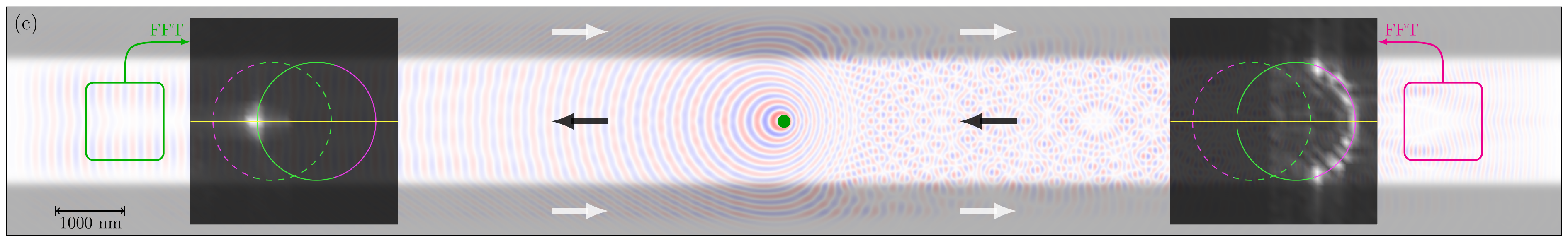}
\caption{The schematic and the magnetic simulations of a spin wave fiber. The black/white arrow denotes the magnetization direction in the each domain colored by yellow/orange (white/gray) for schematic diagram (simulation). (a) The schematic of the spin wave fiber. The magnetization direction in the central core domain is opposite to that in the upper/lower cladding domain. The solid/dashed trajectory is expected for right/left moving spin waves excited at the green bar. The effective magnetic fields in the upper/lower domain walls point in the opposite directions. Consequently, the right moving spin waves are confined in the central core domain, while the left moving modes are refracted into the cladding domains. (b) The simulated propagation of the spin wave beams excited at green bar with exciting frequency $f=100\ \mathrm{GHz}$ (critical angle $\theta_c=52.4^{\circ}$)  and incident angle $\alpha=62.5^\circ$, confirming the expected paths in (a). (c) The simulated propagation of the spin wave excited by a point source with exciting frequency $f=10\ \mathrm{GHz}$. Left/right inset: The Fast Fourier Transformation (FFT) of the spin wave pattern in the region enclosed by the green/magenta box in the far left/right zone. The solid (dashed) circle is for the isofreqency circle of the central (upper/lower) domain, from \Eq{eqn:dispersion}. }
\label{fig:fiber}
\end{figure*}

{\it Spin wave fiber.}
Making use of the total reflection at the interface (domain wall) between two magnetic domains with opposite magnetization directions, a spin wave fiber design is proposed as illustrated in Fig. \ref{fig:fiber}(a). The fiber is consisted of one magnetic domain sandwiched by two magnetic domains with the opposite magnetization direction. Because the effective magnetic fields (directions indicated by the $\otimes$ and $\odot$ symbols in \Figure{fig:fiber}(a)) in the upper and lower domain wall are opposite to each other , the spin wave experiences opposite effective Lorentz forces at  the  upper and lower domain walls. Consequently, spin waves can be transported in the core central domain from left to right by total reflections at both domain walls as indicated by the (magenta) right-going trajectory. In contrast, the spin wave propagating from right to left will be refracted (leaked) into the cladding layers. Therefore, the spin wave fiber is unidirectional.
However, if the spin wave frequency is small enough such that $2\Delta > k_g(\omega)$, the critical angle reaches $\theta_c=-\pi/2$, then total reflections occurs for all spin wave modes in all directions, and the spin wave fiber is bidirectional. A simple estimate of the spin wave coherence length \cite{kajiwara_transmission_2010,cheng_antiferromagnetic_2015} is $l_\phi\sim 1/\alpha_{\text{\tiny G}} k> 100\ \mu\mathrm{m}$, much larger than the typical length of our spin wave fiber structure ($w=\sqrt{A/K}\approx 30 \mathrm{nm}$), ensuring that the functionality of the spin wave fiber is  safe from the damping effect.

The transporting feature of the spin wave fiber is confirmed by magnetic simulations. \Figure{fig:fiber}(b) shows the propagation of a spin wave beam excited at the central core domain with an incident angle $\alpha>\theta_c$. As expected, the right-going spin wave is totally reflected back to the central core domain without leaking to the cladding domains. In contrast, the left-going spin wave passes the domain wall directly and is completely leaked into the upper cladding domain as expected.
\Figure{fig:fiber}(c) shows the spin wave pattern in a spin wave fiber from the point source located at the core domain. The interference pattern of the spin wave on the right side indicates intense reflection of spin wave by the domain walls when propagating rightward, while the simple pattern on the left side indicates the absence of reflection when propagating leftward. By Fourier transforming the spin wave pattern in the highlighted magenta box into $\bk$-space, we see that the spin wave modes in the far right falls onto the magenta arc of the isofrequency circle, which corresponds to the total reflection arc in \Figure{fig:tot_refl}. In contrast, the Fourier transformation of the spin wave in the green box in the far left shows only the direct left moving modes, and all other modes are leaked into the cladding layers by refractions. The exact agreement with the analytical model identifies the critical role of the total reflection in the unidirectionality of the spin wave fiber.

{\it Conclusion.}
In conclusion, in the presence of the Dzyaloshinskii-Moriya interaction, we discovered the generalized Snell's law that governs the spin wave scattering behavior at a magnetic domain wall. Similar to the optical case, a spin wave experiences a total reflection when the incident angle is larger than a critical angle. Using this property, we designed a spin wave fiber that can transport spin waves over long distance, which is confirmed by micromagnetic simulations. The proposed spin wave fiber may be used to interconnect different magnonic computation units.

{\it Methods.} The simulations are performed in COMSOL Multiphysics using the mathematical module where the LLG equation is transformed into weak form and solved by the generalized-alpha method (amplification of high frequency is 0.6). The sample is a YIG thin film with the following parameters: the easy-axis anisotropy $K/\gamma=3.88\times 10^{4}\ \mathrm{A}/\mathrm{m}$,  the exchange constant $A/\gamma=3.28\times 10^{-11}\ \mathrm{A}\cdot \mathrm{m}$, the gyromagnetic ratio $\gamma=  2.21 \times 10^{5}\  \mathrm{rad} \cdot \mathrm{Hz}/(\mathrm{A}/\mathrm{m})$\cite{yan_all-magnonic_2011}, and the DMI constant $D/\gamma=2\times 10^{-3}\ \mathrm{A}$.
The dipolar interaction is neglected for this operating frequency of exchange spin waves.
 The Gaussian spin wave beam is  prepared by introducing a narrow rectangle excitation area, where an oscillating magnetic field is provided, with its amplitude changing in Gaussian form in the longitudinal direction and keeping constant in the transverse direction. \cite{gruszecki_influence_2015}.

 {\it Acknowledgements.} This work was supported by the National Natural Science Foundation of China under Grant No.  11474065, National Basic Research Program of China under Grant No. 2014CB921600 and No. 2015CB921400. R. W. was also supported by the Department of Energy (U. S.) under Grant No. DE-FG02-05ER46237.




\bibliography{refs}

\begin{thebibliography}{27}%
\makeatletter
\providecommand \@ifxundefined [1]{%
 \@ifx{#1\undefined}
}%
\providecommand \@ifnum [1]{%
 \ifnum #1\expandafter \@firstoftwo
 \else \expandafter \@secondoftwo
 \fi
}%
\providecommand \@ifx [1]{%
 \ifx #1\expandafter \@firstoftwo
 \else \expandafter \@secondoftwo
 \fi
}%
\providecommand \natexlab [1]{#1}%
\providecommand \enquote  [1]{``#1''}%
\providecommand \bibnamefont  [1]{#1}%
\providecommand \bibfnamefont [1]{#1}%
\providecommand \citenamefont [1]{#1}%
\providecommand \href@noop [0]{\@secondoftwo}%
\providecommand \href [0]{\begingroup \@sanitize@url \@href}%
\providecommand \@href[1]{\@@startlink{#1}\@@href}%
\providecommand \@@href[1]{\endgroup#1\@@endlink}%
\providecommand \@sanitize@url [0]{\catcode `\\12\catcode `\$12\catcode
  `\&12\catcode `\#12\catcode `\^12\catcode `\_12\catcode `\%12\relax}%
\providecommand \@@startlink[1]{}%
\providecommand \@@endlink[0]{}%
\providecommand \url  [0]{\begingroup\@sanitize@url \@url }%
\providecommand \@url [1]{\endgroup\@href {#1}{\urlprefix }}%
\providecommand \urlprefix  [0]{URL }%
\providecommand \Eprint [0]{\href }%
\providecommand \doibase [0]{http://dx.doi.org/}%
\providecommand \selectlanguage [0]{\@gobble}%
\providecommand \bibinfo  [0]{\@secondoftwo}%
\providecommand \bibfield  [0]{\@secondoftwo}%
\providecommand \translation [1]{[#1]}%
\providecommand \BibitemOpen [0]{}%
\providecommand \bibitemStop [0]{}%
\providecommand \bibitemNoStop [0]{.\EOS\space}%
\providecommand \EOS [0]{\spacefactor3000\relax}%
\providecommand \BibitemShut  [1]{\csname bibitem#1\endcsname}%
\let\auto@bib@innerbib\@empty
\bibitem [{\citenamefont {Kruglyak}\ \emph {et~al.}(2010)\citenamefont
  {Kruglyak}, \citenamefont {Demokritov},\ and\ \citenamefont
  {Grundler}}]{kruglyak_magnonics_2010}%
  \BibitemOpen
  \bibfield  {author} {\bibinfo {author} {\bibfnamefont {V.~V.}\ \bibnamefont
  {Kruglyak}}, \bibinfo {author} {\bibfnamefont {S.~O.}\ \bibnamefont
  {Demokritov}}, \ and\ \bibinfo {author} {\bibfnamefont {D.}~\bibnamefont
  {Grundler}},\ }\href {\doibase 10.1088/0022-3727/43/26/264001} {\bibfield
  {journal} {\bibinfo  {journal} {J. Phys. D: Appl. Phys.}\ }\textbf {\bibinfo
  {volume} {43}},\ \bibinfo {pages} {264001} (\bibinfo {year}
  {2010})}\BibitemShut {NoStop}%
\bibitem [{\citenamefont {Chumak}\ \emph {et~al.}(2015)\citenamefont {Chumak},
  \citenamefont {Vasyuchka}, \citenamefont {Serga},\ and\ \citenamefont
  {Hillebrands}}]{chumak_magnon_2015}%
  \BibitemOpen
  \bibfield  {author} {\bibinfo {author} {\bibfnamefont {A.~V.}\ \bibnamefont
  {Chumak}}, \bibinfo {author} {\bibfnamefont {V.~I.}\ \bibnamefont
  {Vasyuchka}}, \bibinfo {author} {\bibfnamefont {A.~A.}\ \bibnamefont
  {Serga}}, \ and\ \bibinfo {author} {\bibfnamefont {B.}~\bibnamefont
  {Hillebrands}},\ }\href {\doibase 10.1038/nphys3347} {\bibfield  {journal}
  {\bibinfo  {journal} {Nat. Phys.}\ }\textbf {\bibinfo {volume} {11}},\
  \bibinfo {pages} {453} (\bibinfo {year} {2015})}\BibitemShut {NoStop}%
\bibitem [{\citenamefont {Lan}\ \emph {et~al.}(2015)\citenamefont {Lan},
  \citenamefont {Yu}, \citenamefont {Wu},\ and\ \citenamefont
  {Xiao}}]{lan_spin-wave_2015}%
  \BibitemOpen
  \bibfield  {author} {\bibinfo {author} {\bibfnamefont {J.}~\bibnamefont
  {Lan}}, \bibinfo {author} {\bibfnamefont {W.}~\bibnamefont {Yu}}, \bibinfo
  {author} {\bibfnamefont {R.}~\bibnamefont {Wu}}, \ and\ \bibinfo {author}
  {\bibfnamefont {J.}~\bibnamefont {Xiao}},\ }\href {\doibase
  10.1103/PhysRevX.5.041049} {\bibfield  {journal} {\bibinfo  {journal} {Phys.
  Rev. X}\ }\textbf {\bibinfo {volume} {5}},\ \bibinfo {pages} {041049}
  (\bibinfo {year} {2015})}\BibitemShut {NoStop}%
\bibitem [{\citenamefont {Kajiwara}\ \emph {et~al.}(2010)\citenamefont
  {Kajiwara}, \citenamefont {Harii}, \citenamefont {Takahashi}, \citenamefont
  {Ohe}, \citenamefont {Uchida}, \citenamefont {Mizuguchi}, \citenamefont
  {Umezawa}, \citenamefont {Kawai}, \citenamefont {Ando}, \citenamefont
  {Takanashi}, \citenamefont {Maekawa},\ and\ \citenamefont
  {Saitoh}}]{kajiwara_transmission_2010}%
  \BibitemOpen
  \bibfield  {author} {\bibinfo {author} {\bibfnamefont {Y.}~\bibnamefont
  {Kajiwara}}, \bibinfo {author} {\bibfnamefont {K.}~\bibnamefont {Harii}},
  \bibinfo {author} {\bibfnamefont {S.}~\bibnamefont {Takahashi}}, \bibinfo
  {author} {\bibfnamefont {J.}~\bibnamefont {Ohe}}, \bibinfo {author}
  {\bibfnamefont {K.}~\bibnamefont {Uchida}}, \bibinfo {author} {\bibfnamefont
  {M.}~\bibnamefont {Mizuguchi}}, \bibinfo {author} {\bibfnamefont
  {H.}~\bibnamefont {Umezawa}}, \bibinfo {author} {\bibfnamefont
  {H.}~\bibnamefont {Kawai}}, \bibinfo {author} {\bibfnamefont
  {K.}~\bibnamefont {Ando}}, \bibinfo {author} {\bibfnamefont {K.}~\bibnamefont
  {Takanashi}}, \bibinfo {author} {\bibfnamefont {S.}~\bibnamefont {Maekawa}},
  \ and\ \bibinfo {author} {\bibfnamefont {E.}~\bibnamefont {Saitoh}},\ }\href
  {\doibase 10.1038/nature08876} {\bibfield  {journal} {\bibinfo  {journal}
  {Nature}\ }\textbf {\bibinfo {volume} {464}},\ \bibinfo {pages} {262}
  (\bibinfo {year} {2010})}\BibitemShut {NoStop}%
\bibitem [{\citenamefont {Vogt}\ \emph {et~al.}(2014)\citenamefont {Vogt},
  \citenamefont {Fradin}, \citenamefont {Pearson}, \citenamefont {Sebastian},
  \citenamefont {Bader}, \citenamefont {Hillebrands}, \citenamefont
  {Hoffmann},\ and\ \citenamefont {Schultheiss}}]{vogt_realization_2014}%
  \BibitemOpen
  \bibfield  {author} {\bibinfo {author} {\bibfnamefont {K.}~\bibnamefont
  {Vogt}}, \bibinfo {author} {\bibfnamefont {F.}~\bibnamefont {Fradin}},
  \bibinfo {author} {\bibfnamefont {J.}~\bibnamefont {Pearson}}, \bibinfo
  {author} {\bibfnamefont {T.}~\bibnamefont {Sebastian}}, \bibinfo {author}
  {\bibfnamefont {S.}~\bibnamefont {Bader}}, \bibinfo {author} {\bibfnamefont
  {B.}~\bibnamefont {Hillebrands}}, \bibinfo {author} {\bibfnamefont
  {A.}~\bibnamefont {Hoffmann}}, \ and\ \bibinfo {author} {\bibfnamefont
  {H.}~\bibnamefont {Schultheiss}},\ }\href {\doibase 10.1038/ncomms4727}
  {\bibfield  {journal} {\bibinfo  {journal} {Nat. Commun.}\ }\textbf {\bibinfo
  {volume} {5}},\ \bibinfo {pages} {3727} (\bibinfo {year} {2014})}\BibitemShut
  {NoStop}%
\bibitem [{\citenamefont {Chumak}\ \emph {et~al.}(2014)\citenamefont {Chumak},
  \citenamefont {Serga},\ and\ \citenamefont
  {Hillebrands}}]{chumak_magnon_2014}%
  \BibitemOpen
  \bibfield  {author} {\bibinfo {author} {\bibfnamefont {A.~V.}\ \bibnamefont
  {Chumak}}, \bibinfo {author} {\bibfnamefont {A.~A.}\ \bibnamefont {Serga}}, \
  and\ \bibinfo {author} {\bibfnamefont {B.}~\bibnamefont {Hillebrands}},\
  }\href {\doibase 10.1038/ncomms5700} {\bibfield  {journal} {\bibinfo
  {journal} {Nat. Commun.}\ }\textbf {\bibinfo {volume} {5}},\ \bibinfo {pages}
  {4700} (\bibinfo {year} {2014})}\BibitemShut {NoStop}%
\bibitem [{\citenamefont {Garcia-Sanchez}\ \emph {et~al.}(2015)\citenamefont
  {Garcia-Sanchez}, \citenamefont {Borys}, \citenamefont {Soucaille},
  \citenamefont {Adam}, \citenamefont {Stamps},\ and\ \citenamefont
  {Kim}}]{garcia-sanchez_narrow_2015}%
  \BibitemOpen
  \bibfield  {author} {\bibinfo {author} {\bibfnamefont {F.}~\bibnamefont
  {Garcia-Sanchez}}, \bibinfo {author} {\bibfnamefont {P.}~\bibnamefont
  {Borys}}, \bibinfo {author} {\bibfnamefont {R.}~\bibnamefont {Soucaille}},
  \bibinfo {author} {\bibfnamefont {J.-P.}\ \bibnamefont {Adam}}, \bibinfo
  {author} {\bibfnamefont {R.~L.}\ \bibnamefont {Stamps}}, \ and\ \bibinfo
  {author} {\bibfnamefont {J.-V.}\ \bibnamefont {Kim}},\ }\href@noop {}
  {\bibfield  {journal} {\bibinfo  {journal} {Phys. Rev. Lett.}\ }\textbf
  {\bibinfo {volume} {114}},\ \bibinfo {pages} {247206} (\bibinfo {year}
  {2015})}\BibitemShut {NoStop}%
\bibitem [{\citenamefont {Cheng}\ \emph {et~al.}(2015)\citenamefont {Cheng},
  \citenamefont {Daniels}, \citenamefont {Zhu},\ and\ \citenamefont
  {Xiao}}]{cheng_antiferromagnetic_2015}%
  \BibitemOpen
  \bibfield  {author} {\bibinfo {author} {\bibfnamefont {R.}~\bibnamefont
  {Cheng}}, \bibinfo {author} {\bibfnamefont {M.~W.}\ \bibnamefont {Daniels}},
  \bibinfo {author} {\bibfnamefont {J.-G.}\ \bibnamefont {Zhu}}, \ and\
  \bibinfo {author} {\bibfnamefont {D.}~\bibnamefont {Xiao}},\ }\href
  {http://arxiv.org/abs/1509.05295} {\bibfield  {journal} {\bibinfo  {journal}
  {arXiv:1509.05295 [cond-mat]}\ } (\bibinfo {year} {2015})}\BibitemShut
  {NoStop}%
\bibitem [{\citenamefont
  {Dzyaloshinsky}(1958)}]{dzyaloshinsky_thermodynamic_1958}%
  \BibitemOpen
  \bibfield  {author} {\bibinfo {author} {\bibfnamefont {I.}~\bibnamefont
  {Dzyaloshinsky}},\ }\href {\doibase 10.1016/0022-3697(58)90076-3} {\bibfield
  {journal} {\bibinfo  {journal} {J. Phys. Chem. Solids}\ }\textbf {\bibinfo
  {volume} {4}},\ \bibinfo {pages} {241} (\bibinfo {year} {1958})}\BibitemShut
  {NoStop}%
\bibitem [{\citenamefont {Moriya}(1960)}]{moriya_anisotropic_1960}%
  \BibitemOpen
  \bibfield  {author} {\bibinfo {author} {\bibfnamefont {T.}~\bibnamefont
  {Moriya}},\ }\href {\doibase 10.1103/PhysRev.120.91} {\bibfield  {journal}
  {\bibinfo  {journal} {Phys. Rev.}\ }\textbf {\bibinfo {volume} {120}},\
  \bibinfo {pages} {91} (\bibinfo {year} {1960})}\BibitemShut {NoStop}%
\bibitem [{\citenamefont {Senior}\ and\ \citenamefont
  {Jamro}(2009)}]{senior_optical_2009}%
  \BibitemOpen
  \bibfield  {author} {\bibinfo {author} {\bibfnamefont {J.~M.}\ \bibnamefont
  {Senior}}\ and\ \bibinfo {author} {\bibfnamefont {M.~Y.}\ \bibnamefont
  {Jamro}},\ }\href@noop {} {\emph {\bibinfo {title} {Optical fiber
  communications: principles and practice}}}\ (\bibinfo  {publisher} {Pearson
  Education},\ \bibinfo {year} {2009})\BibitemShut {NoStop}%
\bibitem [{\citenamefont {Bak}\ and\ \citenamefont
  {Jensen}(1980)}]{bak_theory_1980}%
  \BibitemOpen
  \bibfield  {author} {\bibinfo {author} {\bibfnamefont {P.}~\bibnamefont
  {Bak}}\ and\ \bibinfo {author} {\bibfnamefont {M.~H.}\ \bibnamefont
  {Jensen}},\ }\href {\doibase 10.1088/0022-3719/13/31/002} {\bibfield
  {journal} {\bibinfo  {journal} {J. Phys. C: Solid State Phys.}\ }\textbf
  {\bibinfo {volume} {13}},\ \bibinfo {pages} {L881} (\bibinfo {year}
  {1980})}\BibitemShut {NoStop}%
\bibitem [{\citenamefont {Moon}\ \emph {et~al.}(2013)\citenamefont {Moon},
  \citenamefont {Seo}, \citenamefont {Lee}, \citenamefont {Kim}, \citenamefont
  {Ryu}, \citenamefont {Lee}, \citenamefont {McMichael},\ and\ \citenamefont
  {Stiles}}]{moon_spin-wave_2013}%
  \BibitemOpen
  \bibfield  {author} {\bibinfo {author} {\bibfnamefont {J.-H.}\ \bibnamefont
  {Moon}}, \bibinfo {author} {\bibfnamefont {S.-M.}\ \bibnamefont {Seo}},
  \bibinfo {author} {\bibfnamefont {K.-J.}\ \bibnamefont {Lee}}, \bibinfo
  {author} {\bibfnamefont {K.-W.}\ \bibnamefont {Kim}}, \bibinfo {author}
  {\bibfnamefont {J.}~\bibnamefont {Ryu}}, \bibinfo {author} {\bibfnamefont
  {H.-W.}\ \bibnamefont {Lee}}, \bibinfo {author} {\bibfnamefont {R.~D.}\
  \bibnamefont {McMichael}}, \ and\ \bibinfo {author} {\bibfnamefont {M.~D.}\
  \bibnamefont {Stiles}},\ }\href {\doibase 10.1103/PhysRevB.88.184404}
  {\bibfield  {journal} {\bibinfo  {journal} {Phys. Rev. B}\ }\textbf {\bibinfo
  {volume} {88}},\ \bibinfo {pages} {184404} (\bibinfo {year}
  {2013})}\BibitemShut {NoStop}%
\bibitem [{\citenamefont {Garcia-Sanchez}\ \emph {et~al.}(2014)\citenamefont
  {Garcia-Sanchez}, \citenamefont {Borys}, \citenamefont {Vansteenkiste},
  \citenamefont {Kim},\ and\ \citenamefont
  {Stamps}}]{garcia_nonreciprocal_2014}%
  \BibitemOpen
  \bibfield  {author} {\bibinfo {author} {\bibfnamefont {F.}~\bibnamefont
  {Garcia-Sanchez}}, \bibinfo {author} {\bibfnamefont {P.}~\bibnamefont
  {Borys}}, \bibinfo {author} {\bibfnamefont {A.}~\bibnamefont
  {Vansteenkiste}}, \bibinfo {author} {\bibfnamefont {J.-V.}\ \bibnamefont
  {Kim}}, \ and\ \bibinfo {author} {\bibfnamefont {R.~L.}\ \bibnamefont
  {Stamps}},\ }\href {\doibase 10.1103/PhysRevB.89.224408} {\bibfield
  {journal} {\bibinfo  {journal} {Phys. Rev. B}\ }\textbf {\bibinfo {volume}
  {89}},\ \bibinfo {pages} {224408} (\bibinfo {year} {2014})}\BibitemShut
  {NoStop}%
\bibitem [{\citenamefont {Yu}\ \emph {et~al.}(2011)\citenamefont {Yu},
  \citenamefont {Genevet}, \citenamefont {Kats}, \citenamefont {Aieta},
  \citenamefont {Tetienne}, \citenamefont {Capasso},\ and\ \citenamefont
  {Gaburro}}]{yu_light_2011}%
  \BibitemOpen
  \bibfield  {author} {\bibinfo {author} {\bibfnamefont {N.}~\bibnamefont
  {Yu}}, \bibinfo {author} {\bibfnamefont {P.}~\bibnamefont {Genevet}},
  \bibinfo {author} {\bibfnamefont {M.~A.}\ \bibnamefont {Kats}}, \bibinfo
  {author} {\bibfnamefont {F.}~\bibnamefont {Aieta}}, \bibinfo {author}
  {\bibfnamefont {J.-P.}\ \bibnamefont {Tetienne}}, \bibinfo {author}
  {\bibfnamefont {F.}~\bibnamefont {Capasso}}, \ and\ \bibinfo {author}
  {\bibfnamefont {Z.}~\bibnamefont {Gaburro}},\ }\href {\doibase
  10.1126/science.1210713} {\bibfield  {journal} {\bibinfo  {journal}
  {Science}\ }\textbf {\bibinfo {volume} {334}},\ \bibinfo {pages} {333}
  (\bibinfo {year} {2011})}\BibitemShut {NoStop}%
\bibitem [{\citenamefont {Yu}\ and\ \citenamefont
  {Capasso}(2014)}]{yu_flat_2014}%
  \BibitemOpen
  \bibfield  {author} {\bibinfo {author} {\bibfnamefont {N.}~\bibnamefont
  {Yu}}\ and\ \bibinfo {author} {\bibfnamefont {F.}~\bibnamefont {Capasso}},\
  }\href {\doibase 10.1038/nmat3839} {\bibfield  {journal} {\bibinfo  {journal}
  {Nat. Mater.}\ }\textbf {\bibinfo {volume} {13}},\ \bibinfo {pages} {139}
  (\bibinfo {year} {2014})}\BibitemShut {NoStop}%
\bibitem [{\citenamefont {Sun}\ \emph {et~al.}(2012{\natexlab{a}})\citenamefont
  {Sun}, \citenamefont {He}, \citenamefont {Xiao}, \citenamefont {Xu},
  \citenamefont {Li},\ and\ \citenamefont {Zhou}}]{sun_gradient_2012}%
  \BibitemOpen
  \bibfield  {author} {\bibinfo {author} {\bibfnamefont {S.}~\bibnamefont
  {Sun}}, \bibinfo {author} {\bibfnamefont {Q.}~\bibnamefont {He}}, \bibinfo
  {author} {\bibfnamefont {S.}~\bibnamefont {Xiao}}, \bibinfo {author}
  {\bibfnamefont {Q.}~\bibnamefont {Xu}}, \bibinfo {author} {\bibfnamefont
  {X.}~\bibnamefont {Li}}, \ and\ \bibinfo {author} {\bibfnamefont
  {L.}~\bibnamefont {Zhou}},\ }\href {\doibase 10.1038/nmat3292} {\bibfield
  {journal} {\bibinfo  {journal} {Nat. Mater.}\ }\textbf {\bibinfo {volume}
  {11}},\ \bibinfo {pages} {426} (\bibinfo {year}
  {2012}{\natexlab{a}})}\BibitemShut {NoStop}%
\bibitem [{\citenamefont {Sun}\ \emph {et~al.}(2012{\natexlab{b}})\citenamefont
  {Sun}, \citenamefont {Yang}, \citenamefont {Wang}, \citenamefont {Juan},
  \citenamefont {Chen}, \citenamefont {Liao}, \citenamefont {He}, \citenamefont
  {Xiao}, \citenamefont {Kung}, \citenamefont {Guo}, \citenamefont {Zhou},\
  and\ \citenamefont {Tsai}}]{sun_high_2012}%
  \BibitemOpen
  \bibfield  {author} {\bibinfo {author} {\bibfnamefont {S.}~\bibnamefont
  {Sun}}, \bibinfo {author} {\bibfnamefont {K.-Y.}\ \bibnamefont {Yang}},
  \bibinfo {author} {\bibfnamefont {C.-M.}\ \bibnamefont {Wang}}, \bibinfo
  {author} {\bibfnamefont {T.-K.}\ \bibnamefont {Juan}}, \bibinfo {author}
  {\bibfnamefont {W.~T.}\ \bibnamefont {Chen}}, \bibinfo {author}
  {\bibfnamefont {C.~Y.}\ \bibnamefont {Liao}}, \bibinfo {author}
  {\bibfnamefont {Q.}~\bibnamefont {He}}, \bibinfo {author} {\bibfnamefont
  {S.}~\bibnamefont {Xiao}}, \bibinfo {author} {\bibfnamefont {W.-T.}\
  \bibnamefont {Kung}}, \bibinfo {author} {\bibfnamefont {G.-Y.}\ \bibnamefont
  {Guo}}, \bibinfo {author} {\bibfnamefont {L.}~\bibnamefont {Zhou}}, \ and\
  \bibinfo {author} {\bibfnamefont {D.~P.}\ \bibnamefont {Tsai}},\ }\href
  {\doibase 10.1021/nl3032668} {\bibfield  {journal} {\bibinfo  {journal} {Nano
  Lett.}\ }\textbf {\bibinfo {volume} {12}},\ \bibinfo {pages} {6223} (\bibinfo
  {year} {2012}{\natexlab{b}})}\BibitemShut {NoStop}%
\bibitem [{\citenamefont {Chen}\ \emph {et~al.}(2012)\citenamefont {Chen},
  \citenamefont {Huang}, \citenamefont {M\"uhlenbernd}, \citenamefont {Li},
  \citenamefont {Bai}, \citenamefont {Tan}, \citenamefont {Jin}, \citenamefont
  {Qiu}, \citenamefont {Zhang},\ and\ \citenamefont
  {Zentgraf}}]{chen_dual-polarity_2012}%
  \BibitemOpen
  \bibfield  {author} {\bibinfo {author} {\bibfnamefont {X.}~\bibnamefont
  {Chen}}, \bibinfo {author} {\bibfnamefont {L.}~\bibnamefont {Huang}},
  \bibinfo {author} {\bibfnamefont {H.}~\bibnamefont {M\"uhlenbernd}}, \bibinfo
  {author} {\bibfnamefont {G.}~\bibnamefont {Li}}, \bibinfo {author}
  {\bibfnamefont {B.}~\bibnamefont {Bai}}, \bibinfo {author} {\bibfnamefont
  {Q.}~\bibnamefont {Tan}}, \bibinfo {author} {\bibfnamefont {G.}~\bibnamefont
  {Jin}}, \bibinfo {author} {\bibfnamefont {C.-W.}\ \bibnamefont {Qiu}},
  \bibinfo {author} {\bibfnamefont {S.}~\bibnamefont {Zhang}}, \ and\ \bibinfo
  {author} {\bibfnamefont {T.}~\bibnamefont {Zentgraf}},\ }\href {\doibase
  10.1038/ncomms2207} {\bibfield  {journal} {\bibinfo  {journal} {Nat.
  Commun.}\ }\textbf {\bibinfo {volume} {3}},\ \bibinfo {pages} {1198}
  (\bibinfo {year} {2012})}\BibitemShut {NoStop}%
\bibitem [{\citenamefont {Li}\ \emph {et~al.}(2014)\citenamefont {Li},
  \citenamefont {Jiang}, \citenamefont {Li}, \citenamefont {Liang},
  \citenamefont {Zou}, \citenamefont {Yin},\ and\ \citenamefont
  {Cheng}}]{li_experimental_2014}%
  \BibitemOpen
  \bibfield  {author} {\bibinfo {author} {\bibfnamefont {Y.}~\bibnamefont
  {Li}}, \bibinfo {author} {\bibfnamefont {X.}~\bibnamefont {Jiang}}, \bibinfo
  {author} {\bibfnamefont {R.-q.}\ \bibnamefont {Li}}, \bibinfo {author}
  {\bibfnamefont {B.}~\bibnamefont {Liang}}, \bibinfo {author} {\bibfnamefont
  {X.-y.}\ \bibnamefont {Zou}}, \bibinfo {author} {\bibfnamefont {L.-l.}\
  \bibnamefont {Yin}}, \ and\ \bibinfo {author} {\bibfnamefont {J.-c.}\
  \bibnamefont {Cheng}},\ }\href {\doibase 10.1103/PhysRevApplied.2.064002}
  {\bibfield  {journal} {\bibinfo  {journal} {Phys. Rev. Appl.}\ }\textbf
  {\bibinfo {volume} {2}},\ \bibinfo {pages} {064002} (\bibinfo {year}
  {2014})}\BibitemShut {NoStop}%
\bibitem [{\citenamefont {Zhu}\ \emph {et~al.}(2015)\citenamefont {Zhu},
  \citenamefont {Zou}, \citenamefont {Liang},\ and\ \citenamefont
  {Cheng}}]{zhu_acoustic_2015}%
  \BibitemOpen
  \bibfield  {author} {\bibinfo {author} {\bibfnamefont {Y.-F.}\ \bibnamefont
  {Zhu}}, \bibinfo {author} {\bibfnamefont {X.-Y.}\ \bibnamefont {Zou}},
  \bibinfo {author} {\bibfnamefont {B.}~\bibnamefont {Liang}}, \ and\ \bibinfo
  {author} {\bibfnamefont {J.-C.}\ \bibnamefont {Cheng}},\ }\href {\doibase
  10.1063/1.4930300} {\bibfield  {journal} {\bibinfo  {journal} {Appl. Phys.
  Lett.}\ }\textbf {\bibinfo {volume} {107}},\ \bibinfo {pages} {113501}
  (\bibinfo {year} {2015})}\BibitemShut {NoStop}%
\bibitem [{\citenamefont {Vashkovskii}\ and\ \citenamefont
  {Lokk}(2004)}]{vashkovskii_negative_2004}%
  \BibitemOpen
  \bibfield  {author} {\bibinfo {author} {\bibfnamefont {A.~V.}\ \bibnamefont
  {Vashkovskii}}\ and\ \bibinfo {author} {\bibfnamefont {E.~H.}\ \bibnamefont
  {Lokk}},\ }\href {\doibase 10.1070/PU2004v047n06ABEH001793} {\bibfield
  {journal} {\bibinfo  {journal} {Phys. Usp.}\ }\textbf {\bibinfo {volume}
  {47}},\ \bibinfo {pages} {601} (\bibinfo {year} {2004})}\BibitemShut
  {NoStop}%
\bibitem [{\citenamefont {Lock}(2008)}]{lock_properties_2008}%
  \BibitemOpen
  \bibfield  {author} {\bibinfo {author} {\bibfnamefont {E.~H.}\ \bibnamefont
  {Lock}},\ }\href {\doibase 10.1070/PU2008v051n04ABEH006460} {\bibfield
  {journal} {\bibinfo  {journal} {Phys. Usp.}\ }\textbf {\bibinfo {volume}
  {51}},\ \bibinfo {pages} {375} (\bibinfo {year} {2008})}\BibitemShut
  {NoStop}%
\bibitem [{\citenamefont {Kim}\ \emph {et~al.}(2008)\citenamefont {Kim},
  \citenamefont {Choi}, \citenamefont {Lee}, \citenamefont {Han}, \citenamefont
  {Jung},\ and\ \citenamefont {Choi}}]{kim_negative_2008}%
  \BibitemOpen
  \bibfield  {author} {\bibinfo {author} {\bibfnamefont {S.-K.}\ \bibnamefont
  {Kim}}, \bibinfo {author} {\bibfnamefont {S.}~\bibnamefont {Choi}}, \bibinfo
  {author} {\bibfnamefont {K.-S.}\ \bibnamefont {Lee}}, \bibinfo {author}
  {\bibfnamefont {D.-S.}\ \bibnamefont {Han}}, \bibinfo {author} {\bibfnamefont
  {D.-E.}\ \bibnamefont {Jung}}, \ and\ \bibinfo {author} {\bibfnamefont
  {Y.-S.}\ \bibnamefont {Choi}},\ }\href {\doibase 10.1063/1.2936294}
  {\bibfield  {journal} {\bibinfo  {journal} {Appl. Phys. Lett.}\ }\textbf
  {\bibinfo {volume} {92}},\ \bibinfo {pages} {212501} (\bibinfo {year}
  {2008})}\BibitemShut {NoStop}%
\bibitem [{\citenamefont {Yan}\ \emph {et~al.}(2011)\citenamefont {Yan},
  \citenamefont {Wang},\ and\ \citenamefont {Wang}}]{yan_all-magnonic_2011}%
  \BibitemOpen
  \bibfield  {author} {\bibinfo {author} {\bibfnamefont {P.}~\bibnamefont
  {Yan}}, \bibinfo {author} {\bibfnamefont {X.~S.}\ \bibnamefont {Wang}}, \
  and\ \bibinfo {author} {\bibfnamefont {X.~R.}\ \bibnamefont {Wang}},\ }\href
  {\doibase 10.1103/PhysRevLett.107.177207} {\bibfield  {journal} {\bibinfo
  {journal} {Phys. Rev. Lett.}\ }\textbf {\bibinfo {volume} {107}},\ \bibinfo
  {pages} {177207} (\bibinfo {year} {2011})}\BibitemShut {NoStop}%
\bibitem [{\citenamefont {Wang}\ \emph {et~al.}(2015)\citenamefont {Wang},
  \citenamefont {Albert}, \citenamefont {Beg}, \citenamefont {Bisotti},
  \citenamefont {Chernyshenko}, \citenamefont {Cortes-Ortuno}, \citenamefont
  {Hawke},\ and\ \citenamefont {Fangohr}}]{wang_magnon-driven_2015}%
  \BibitemOpen
  \bibfield  {author} {\bibinfo {author} {\bibfnamefont {W.}~\bibnamefont
  {Wang}}, \bibinfo {author} {\bibfnamefont {M.}~\bibnamefont {Albert}},
  \bibinfo {author} {\bibfnamefont {M.}~\bibnamefont {Beg}}, \bibinfo {author}
  {\bibfnamefont {M.-A.}\ \bibnamefont {Bisotti}}, \bibinfo {author}
  {\bibfnamefont {D.}~\bibnamefont {Chernyshenko}}, \bibinfo {author}
  {\bibfnamefont {D.}~\bibnamefont {Cortes-Ortuno}}, \bibinfo {author}
  {\bibfnamefont {I.}~\bibnamefont {Hawke}}, \ and\ \bibinfo {author}
  {\bibfnamefont {H.}~\bibnamefont {Fangohr}},\ }\href {\doibase
  10.1103/PhysRevLett.114.087203} {\bibfield  {journal} {\bibinfo  {journal}
  {Phys. Rev. Lett.}\ }\textbf {\bibinfo {volume} {114}},\ \bibinfo {pages}
  {087203} (\bibinfo {year} {2015})}\BibitemShut {NoStop}%
\bibitem [{\citenamefont {Gruszecki}\ \emph {et~al.}(2015)\citenamefont
  {Gruszecki}, \citenamefont {Dadoenkova}, \citenamefont {Dadoenkova},
  \citenamefont {Lyubchanskii}, \citenamefont {Romero-Vivas}, \citenamefont
  {Guslienko},\ and\ \citenamefont {Krawczyk}}]{gruszecki_influence_2015}%
  \BibitemOpen
  \bibfield  {author} {\bibinfo {author} {\bibfnamefont {P.}~\bibnamefont
  {Gruszecki}}, \bibinfo {author} {\bibfnamefont {Y.~S.}\ \bibnamefont
  {Dadoenkova}}, \bibinfo {author} {\bibfnamefont {N.~N.}\ \bibnamefont
  {Dadoenkova}}, \bibinfo {author} {\bibfnamefont {I.~L.}\ \bibnamefont
  {Lyubchanskii}}, \bibinfo {author} {\bibfnamefont {J.}~\bibnamefont
  {Romero-Vivas}}, \bibinfo {author} {\bibfnamefont {K.~Y.}\ \bibnamefont
  {Guslienko}}, \ and\ \bibinfo {author} {\bibfnamefont {M.}~\bibnamefont
  {Krawczyk}},\ }\href {\doibase 10.1103/PhysRevB.92.054427} {\bibfield
  {journal} {\bibinfo  {journal} {Phys. Rev. B}\ }\textbf {\bibinfo {volume}
  {92}},\ \bibinfo {pages} {054427} (\bibinfo {year} {2015})}\BibitemShut
  {NoStop}%
\end{thebibliography}%

\end{document}